# Number of Wavevectors for Each Frequency in a Periodic Structure

**Farhad Farzbod**
Assistant Professor, University of Mississippi
201A Carrier Hall, University, MS 38677
farzbod@olemiss.edu
ASME Member

**ABSTRACT**

*Periodic structures have interesting acoustic and vibration properties making them suitable for a wide variety of applications. In a periodic structure, the number of frequencies for each wavevector depends on the degree of freedom of the unit cell. In this paper, we investigate the number of wavevectors for each frequency. This analysis defines the upper bound for the maximum number of wavevectors for each frequency in a general periodic structure which might include damping. Investigation presented in this paper can also provide an insight for designing materials in which the interaction between unit cells is not limited to the closest neighbor. As an example application of this work, we investigate phonon dispersion curves in hexagonal form of Boron Nitride to show that first neighbor interaction is not sufficient to model dispersion curves with force-constant-model.*

## 1. INTRODUCTION

There has been abundant interest in acoustic and vibrational properties of periodic systems. In most of these analysis Bloch's theorem [1] is employed. Mead et al. [2, 3] used a variant of this theorem to investigate harmonic wave propagation in periodic structures. Other researchers since then applied several variations of the Bloch's theorem in a wide range of applications [4-22]. This includes applications in designing filters [1-3], wave guides [4], wave beaming [5, 6] and nonlinear materials [17, 19, 22]. For a comprehensive review of this field and its future outlook refer to the paper by Hussein et al. [16]. In these works, the nearest neighbor interaction is considered. Because in planar structures, such as honeycomb plates, each unit cell is connected to its closest neighbors only. This is not the case for atoms in a crystal. Each atom is interacting with atoms beyond its closest neighbor. In the next section, we will overview phonon dispersion curves for crystals and then in the following section we will investigate how this neighbors' interaction is related to the number of wave vectors at each frequency.

Analysis presented in this paper gives an insight for designing materials in which not just the closest neighbor is interacting with the unit cell. This can be engineered in structures in which magnets are used, masses have electric charges, or structures in which the links are much lighter than the central masses so links' stiffness enter the equations.

## 2. INTERACTION BEYOND THE NEAREST NEIGHBOR

The clear example for these types of interactions is the interatomic ones in a crystal. In the investigation of phonon dispersion curves of crystals, there have been different approaches such as semiempirical, and *ab initio*, or first-principle calculation. In semiempirical approaches, the model relies on experimental characterization of force field parameters. Phonon dispersion curves are experimentally obtained for some values of wavevectors by means of neutron scattering [23, 24], Raman spectroscopy [25] or Electron energy loss spectroscopy [26]. Interaction potentials are then assumed in a functional form such as Lennard-Jones (LJ) or Morse potential [27]. The constants of these functions are obtained by fitting the experimental results. Since in the semiempirical approach one has to rely on experiments to determine the coefficients of fitting functions, the results are inherently *ad hoc,* and experiments have to be performed for each material. In contrast, when we use the *ab initio* method, all the calculations are based on the laws of physics, such as quantum mechanics and density functional theory [28]. Although the *ab initio* method is more general, it is a computationally costly method, and in most cases it is overly complicated.

In the semiempirical approach, the potential energy function resembles a quadratic function when particles are close to the equilibrium position. Hence, in the semiempirical calculation of phonon dispersion curves, the potential energy between the particles assumed to be quadratic. In this kind of model - called force constant model - forces are represented by linear springs and then the spring constants are found by fitting the theoretical phonon dispersion curves with the experimental dispersion curves.

Generally speaking, in a semiempirical method, it is assumed that particles separated beyond certain cut-off distances, have negligible force effect, and hence it is safe to neglect them in the calculation. It is however, the matter of best fitting that dictates the cut-off distance. Jishi et al. [29] showed that to calculate phonon dispersion curves for graphite, it suffices to consider the interaction of atoms up to the fourth nearest neighbor, both interplane and intraplane. A model with less than a fourth nearest neighbor cannot reproduce the experimental results. The same fourth nearest neighbor model has been shown [30-32] to agree well with the experimental results for graphene, C-60 and nanotubules. Currently, the number of neighbor interactions needed for accurate fitting is a matter of try and error. In this paper, we lay a foundation to calculate the necessary minimum number of neighbors needed in a model based on the number of wave vectors for each frequency. This theory cannot show the sufficient number, nevertheless, it can be considered as a tool helping the modeling process.

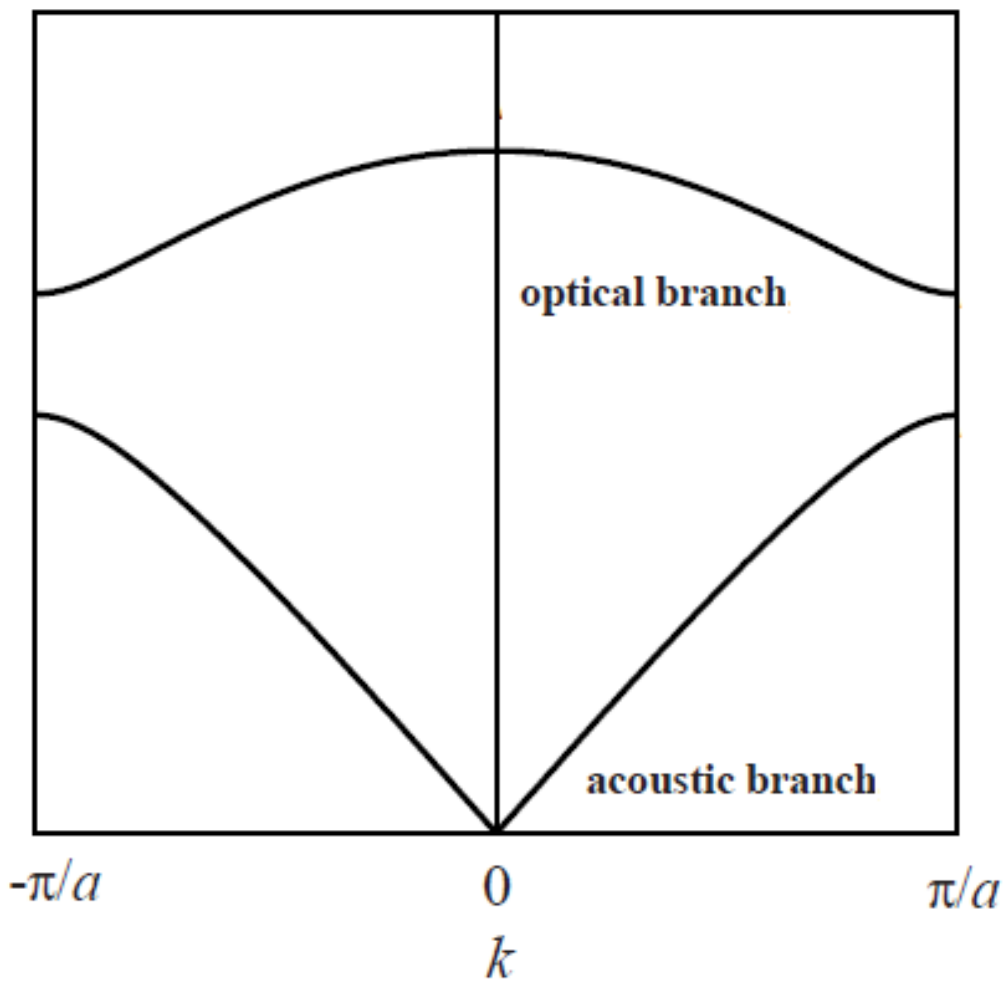

*Figure 1 Dispersion curve for a general diatomic chain*

## 3. THEORY

In structures with no energy dissipation, the number of temporal frequencies ω for each wavevector depends on the degree of freedom of a unit cell. In other words, the number of dispersion branches equates to the degree of freedom of the unit cell.

In this section we investigate the number of wavevectors for each ω in a dispersion relation. We focus on discrete structures; any unit cell of a periodic structure can be modelled by a discrete system of equations after invoking finite element analysis. In a bimaterial chain (or diatomic crystal chain), if we consider only the nearest neighbor interaction, the optical and acoustic branches are monotonic; meaning that for each ω there would be only one corresponding positive wavevector in the first Brillouin zone [33] as is evident in the dispersion curve for an example diatomic chain in Fig. 1. If we increase the number of interactions or consider two-dimensional lattices, there might be more than one wavevector corresponding to each ω. This fact was first investigated in a rigorous way by Brillouin [1] for a simple monatomic chain. In this section, we investigate this for structures with two or three dimensions.

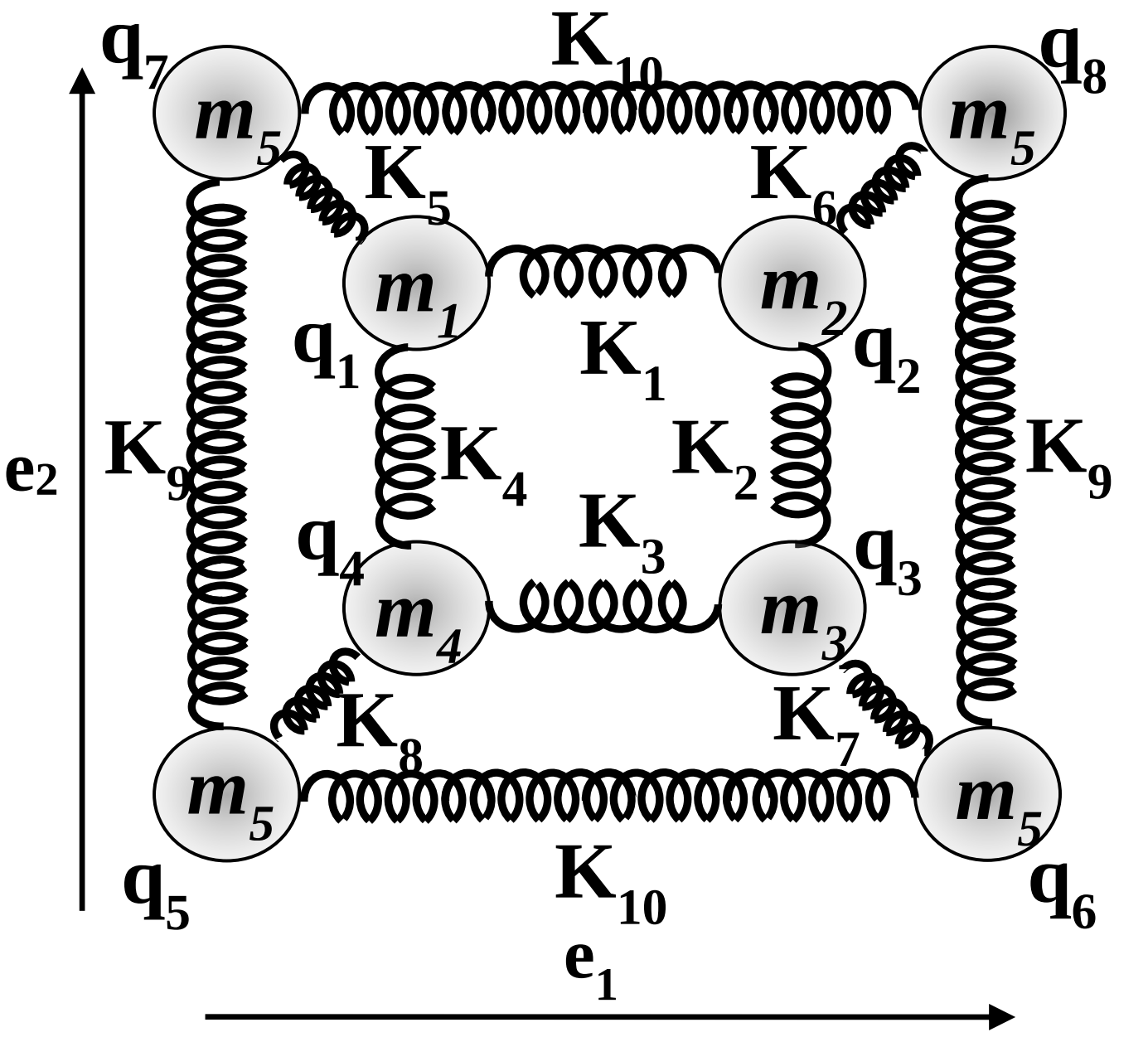

Figure 2. A sample 2D structure with out of plane motion and some internal masses.

We follow our previous technique [34-36] of defining a minimum set of displacement for a unit cell and a set of linear operators as:

$$\mathbf{q}=\begin{bmatrix}\mathbf{q_i}\\ \tilde{\mathbf{q}}\\ \tilde{\mathbf{q}}_\mathbf{x}\\ \tilde{\mathbf{q}}_\mathbf{y}\\ \tilde{\mathbf{q}}_\mathbf{xy}\end{bmatrix}=\begin{bmatrix}\mathbf{I} & \mathbf{0}\\ \mathbf{0} & \mathbf{I}\\ \mathbf{0} & \mathbf{T_x}\\ \mathbf{0} & \mathbf{T_y}\\ \mathbf{0} & \mathbf{T_{xy}}\end{bmatrix}\begin{bmatrix}\mathbf{q_i}\\ \tilde{\mathbf{q}}\end{bmatrix}=\mathbf{T}\begin{bmatrix}\mathbf{q_i}\\ \tilde{\mathbf{q}}\end{bmatrix}=\mathbf{T}\hat{\mathbf{q}} \qquad (1)$$

Where $\boldsymbol{q_i}$ and $\tilde{\boldsymbol{q}}$ are internal and essential boundary displacements, respectively. In eq. (1), $\tilde{\boldsymbol{q}}_{\boldsymbol{x}}=\boldsymbol{T_x}\tilde{\boldsymbol{q}}$, $\tilde{\boldsymbol{q}}_{\boldsymbol{y}}=\boldsymbol{T_y}\tilde{\boldsymbol{q}}$, $\tilde{\boldsymbol{q}}_{\boldsymbol{xy}}=\boldsymbol{T_{xy}}\tilde{\boldsymbol{q}}$ and $\boldsymbol{I}$'s are in proper dimensions. Linear operator $\boldsymbol{T_x}$, $\boldsymbol{T_y}$ and $\boldsymbol{T_{xy}}$ are push-forward operators in $e_1$ and $e_2$ direction which are along the unit vectors of the structure (not necessary orthogonal). They contain $e^{2\pi i k_1}$ and $e^{2\pi i k_2}$ where $k_1$ and $k_2$ are complex wavevectors. For the sake of simplicity we call them $x$ and $y$. For the example structure of Fig. 2 we have:

$$\mathbf{q_i} = \begin{bmatrix} q_1 \\ q_2 \\ q_3 \\ q_4 \end{bmatrix}, \tilde{\mathbf{q}} = [q_5], \tilde{\mathbf{q}}_\mathbf{x} = [q_6], \tilde{\mathbf{q}}_\mathbf{y} = [q_7], \tilde{\mathbf{q}}_\mathbf{xy} = [q_8] \tag{2}$$

Similarly for the example structure of Fig. 3 we have:

$$\mathbf{q_i} = \begin{bmatrix} q_1 \\ q_2 \end{bmatrix}, \tilde{\mathbf{q}} = \begin{bmatrix} q_3 \\ q_4 \end{bmatrix}, \tilde{\mathbf{q}}_\mathbf{x} = [q_5], \tilde{\mathbf{q}}_\mathbf{y} = [q_6], \tilde{\mathbf{q}}_\mathbf{xy} = \begin{bmatrix} q_7 \\ q_8 \end{bmatrix} \tag{3}$$

The equation of motion for a unit cell takes the form of $(-\omega^2\mathbf{M} + i\omega\mathbf{C} + \mathbf{K})\mathbf{T}\hat{\mathbf{q}} = \mathbf{F}.$ In order to cancel out the force term on the right, we multiply both sides by the pull-back operator $\overline{\mathbf{T}}^T$ defined as:

$$\overline{\mathbf{T}}^T = \begin{bmatrix} \mathbf{I} & \mathbf{0} & \mathbf{0} & \mathbf{0} & \mathbf{0} \\ \mathbf{0} & \mathbf{I} & \mathbf{T}_{-x}^T & \mathbf{T}_{-y}^T & \mathbf{T}_{-xy}^T \end{bmatrix} \tag{4}$$

The dispersion relation is obtained by setting the determinant of the resultant equation to zero:

$$\det(\overline{\mathbf{T}}^T(-\omega^2\mathbf{M} + i\omega\mathbf{C} + \mathbf{K})\mathbf{T}) = \mathbf{0} \tag{5}$$

In our previous work we partitioned displacements. In this section we partition mass, stiffness and damping matrices accordingly; the mass and stiffness matrices are expressed as:

$$\mathbf{M} = \begin{bmatrix} \mathbf{M}_i & \mathbf{0} & \mathbf{0} & \mathbf{0} & \mathbf{0} \\ \mathbf{0} & \mathbf{M}_m & \mathbf{0} & \mathbf{0} & \mathbf{0} \\ \mathbf{0} & \mathbf{0} & \mathbf{M}_x & \mathbf{0} & \mathbf{0} \\ \mathbf{0} & \mathbf{0} & \mathbf{0} & \mathbf{M}_y & \mathbf{0} \\ \mathbf{0} & \mathbf{0} & \mathbf{0} & \mathbf{0} & \mathbf{M}_{xy} \end{bmatrix}, \mathbf{K} = \begin{bmatrix} \mathbf{K}_i & \mathbf{K}_{im} & \mathbf{K}_{ix} & \mathbf{K}_{iy} & \mathbf{K}_{ixy} \\ \mathbf{K}_{im}^T & \mathbf{K}_m & \mathbf{K}_{mx} & \mathbf{K}_{my} & \mathbf{K}_{mxy} \\ \mathbf{K}_{ix}^T & \mathbf{K}_{mx}^T & \mathbf{K}_x & \mathbf{K}_{xy} & \mathbf{K}_{xxy} \\ \mathbf{K}_{iy}^T & \mathbf{K}_{my}^T & \mathbf{K}_{xy}^T & \mathbf{K}_y & \mathbf{K}_{yxy} \\ \mathbf{K}_{ixy}^T & \mathbf{K}_{mxy}^T & \mathbf{K}_{xxy}^T & \mathbf{K}_{yxy}^T & \mathbf{K}_{xyxy} \end{bmatrix} \tag{6}$$

Damping matrix is partitioned in the same fashion as the stiffness matrix. Note that the elements of the mass, stiffness and damping matrices are matrices themselves properly dimensioned. For example, if $\mathbf{q_i}$ and $\tilde{\mathbf{q}}_\mathbf{x}$ are *n×1* and *r×1*, respectively, then $\mathbf{K}_{ix}$ is *n x r*.

The number of wavevectors for each ω depends on the degree of *x* and *y* in the polynomial of eq. (5). As stated before *x* and *y* represent $e^{2\pi i k_1}$ and $e^{2\pi i k_2}$ in $\mathbf{T}_x$, $\mathbf{T}_y$ and $\mathbf{T}_{xy}$. Meanwhile $\mathbf{T}_{-\mathbf{x}}^T$, $\mathbf{T}_{-\mathbf{y}}^T$ and $\mathbf{T}_{-\mathbf{xy}}^T$ are pull-back operators containing $e^{-2\pi i k_1}$ and $e^{-2\pi i k_2}$which can be replaced by $x^{-1}$ and $y^{-1}$. In order to investigate the degree of *x* and *y* in the polynomial, we first decompose $\mathbf{T}$ and $\overline{\mathbf{T}}^T$ into terms with and without these variables. Note that all the elements in $\mathbf{T}_x$ have *x*; there is no term with higher powers of *x* and no term without *x*. As a result, we can factor out *x* from$\mathbf{T}_x$. By the same argument, we can factor out *y*, *xy*, $x^{-1}$, $y^{-1}$ and $x^{-1}y^{-1}$ from $\mathbf{T_y}$ ,$\mathbf{T_{xy}}$ ,$\mathbf{T}_{-x}^T$ ,$\mathbf{T}_{-y}^T$ and $\mathbf{T}_{-xy}^T$ , respectively. We then rephrase these matrices as:

$$\mathbf{T}_{\boldsymbol{x}} = x\widehat{\mathbf{T}}_{\boldsymbol{x}}\,, \qquad \mathbf{T}_{-\boldsymbol{x}}^T = x^{-1}\widehat{\mathbf{T}}_{-x}^T$$
$$\mathbf{T}_{\boldsymbol{y}} = y\widehat{\mathbf{T}}_{\boldsymbol{x}}\,, \qquad \mathbf{T}_{-\boldsymbol{y}}^T = y^{-1}\widehat{\mathbf{T}}_{-y}^T \tag{7}$$
$$\mathbf{T}_{\boldsymbol{xy}} = xy\widehat{\mathbf{T}}_{\boldsymbol{xy}}\,, \qquad \mathbf{T}_{-\boldsymbol{xy}}^T = x^{-1}y^{-1}\widehat{\mathbf{T}}_{-x}^T$$

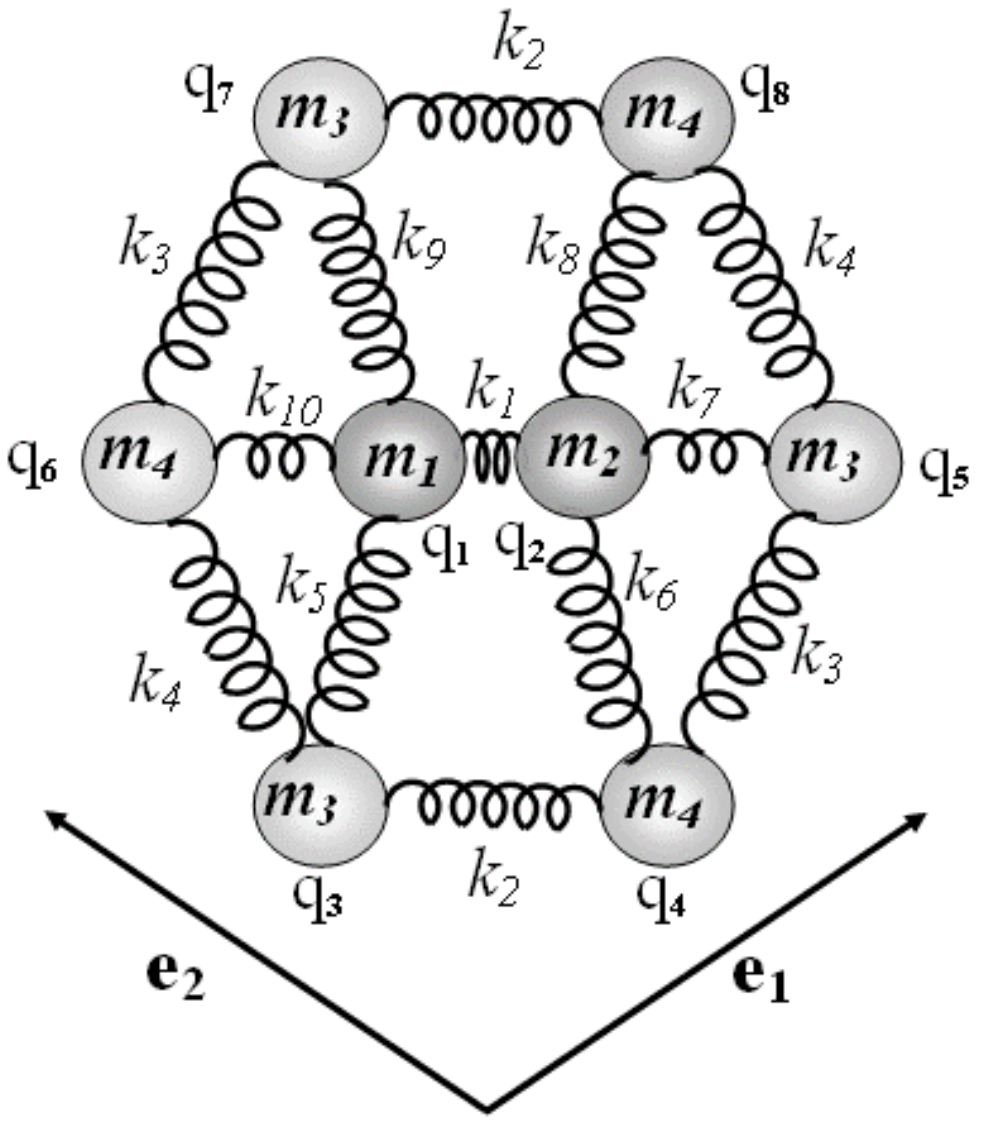

Figure 3: 2D structure with out of plane motion

In the example structure of Fig. 3, these linear transformations take the form of:

$$\mathbf{T}_{\boldsymbol{x}} = [\mathbf{I}e^{2\pi i k_1} \quad \mathbf{0}] = x[\mathbf{I} \quad \mathbf{0}] = x\widehat{\mathbf{T}}_{\boldsymbol{x}},\ \mathbf{T}_{\boldsymbol{y}} = [\mathbf{0} \quad \mathbf{I}e^{2\pi i k_2}] = y[\mathbf{0} \quad \mathbf{I}] = x\widehat{\mathbf{T}}_{\boldsymbol{y}}$$
$$\mathbf{T}_{-\boldsymbol{x}}^T = \begin{bmatrix}\mathbf{I}e^{-2\pi i k_1}\\ \mathbf{0}\end{bmatrix} = x^{-1}\begin{bmatrix}\mathbf{I}\\ \mathbf{0}\end{bmatrix} = x^{-1}\widehat{\mathbf{T}}_{-x}^T,\ \mathbf{T}_{-\boldsymbol{y}}^T = \begin{bmatrix}\mathbf{0}\\ \mathbf{I}e^{-2\pi i k_2}\end{bmatrix} = y^{-1}\begin{bmatrix}\mathbf{0}\\ \mathbf{I}\end{bmatrix} = y^{-1}\widehat{\mathbf{T}}_{-y}^T$$
$$\mathbf{T}_{\boldsymbol{xy}} = \begin{bmatrix}\mathbf{I}e^{2\pi i(k_1+k_2)} & \mathbf{0}\\ \mathbf{0} & \mathbf{I}e^{2\pi i(k_1+k_2)}\end{bmatrix} = xy\begin{bmatrix}\mathbf{I} & \mathbf{0}\\ \mathbf{0} & \mathbf{I}\end{bmatrix} = xy\widehat{\mathbf{T}}_{\boldsymbol{xy}} \tag{8}$$
$$\mathbf{T}_{-xy}^T = \begin{bmatrix}\mathbf{I}e^{-2\pi i(k_1+k_2)} & \mathbf{0}\\ \mathbf{0} & \mathbf{I}e^{-2\pi i(k_1+k_2)}\end{bmatrix} = x^{-1}y^{-1}\begin{bmatrix}\mathbf{I} & \mathbf{0}\\ \mathbf{0} & \mathbf{I}\end{bmatrix} = x^{-1}y^{-1}\widehat{\mathbf{T}}_{-xy}^T$$

Since the pulling-back and pushing-forward operations of the linear operators in the $\boldsymbol{e_1}$ direction are embedded in *x* and $x^{-1}$, it can be easily verified that:

$$\widehat{\mathbf{T}}_x^T = \widehat{\mathbf{T}}_{-x}^T \tag{9}$$

Similarly,

$$\widehat{\mathbf{T}}_y^T = \widehat{\mathbf{T}}_{-y}^T\,,\ \widehat{\mathbf{T}}_{xy}^T = \widehat{\mathbf{T}}_{-xy}^T \tag{10}$$

Let $\mathbf{D} = \overline{\mathbf{T}}^T(-\omega^2\mathbf{M} + i\omega\mathbf{C} + \mathbf{K})\mathbf{T}$ , then by substituting partitioned **M**, **C** and **K** from eq. (6) and using elements $\overline{\mathbf{T}}^T$ and $\mathbf{T}$ in the form of eq. (8), we get:

$$\mathbf{D} = \begin{bmatrix}\mathbf{D}_{11} & \mathbf{D}_{12}\\ \mathbf{D}_{21} & \mathbf{D}_{22}\end{bmatrix}, \tag{11}$$

in which,

$$\mathbf{D}_{11} = -\omega^2\,\mathbf{M}_i\, + i\omega\mathbf{C}_i + \mathbf{K}_i \tag{12}$$

$$\mathbf{D}_{12} = i\omega\mathbf{C}_{im} + \mathbf{K}_{im} + (i\omega\mathbf{C}_{ix} + \mathbf{K}_{ix})x\widehat{\mathbf{T}}_x + \big(i\omega\mathbf{C}_{iy} + \mathbf{K}_{iy}\big)y\widehat{\mathbf{T}}_y + \big(i\omega\mathbf{C}_{ixy} + \mathbf{K}_{ixy}\big)xy\widehat{\mathbf{T}}_{xy} \tag{13}$$

$$\mathbf{D}_{21} = i\omega\mathbf{C}_{im}^T + \mathbf{K}_{im}^T + \frac{\widehat{\mathbf{T}}_x^T(i\omega\mathbf{C}_{ix}^T + \mathbf{K}_{ix}^T)}{x} + \frac{\widehat{\mathbf{T}}_y^T(i\omega\mathbf{C}_{iy}^T + \mathbf{K}_{iy}^T)}{y} + \frac{\widehat{\mathbf{T}}_{xy}^T(i\omega\mathbf{C}_{ixy}^T + \mathbf{K}_{ixy}^T)}{xy} \tag{14}$$

$$
\begin{aligned}
\mathbf{D}_{22} = -\omega^2\,\mathbf{M}_m &+ i\omega\mathbf{C}_m + \mathbf{K}_m + (i\omega\mathbf{C}_{mx} + \mathbf{K}_{mx})x\widehat{\mathbf{T}}_x \\
&+ \left(i\omega\mathbf{C}_{my} + \mathbf{K}_{my}\right)y\widehat{\mathbf{T}}_y + \left(i\omega\mathbf{C}_{mxy} + \mathbf{K}_{mxy}\right)xy\widehat{\mathbf{T}}_{xy} \\
&+ \widehat{\mathbf{T}}_x^T(-\omega^2\,\mathbf{M}_x + i\omega\mathbf{C}_x + \mathbf{K}_x)\widehat{\mathbf{T}}_x + \widehat{\mathbf{T}}_x^T\left(i\omega\mathbf{C}_{xxy} + \mathbf{K}_{xxy}\right)y\widehat{\mathbf{T}}_{xy} \\
&+ \frac{\widehat{\mathbf{T}}_x^T(i\omega\mathbf{C}_{mx}^T + \mathbf{K}_{mx}^T)}{x} + \frac{\widehat{\mathbf{T}}_x^T\left(i\omega\mathbf{C}_{xy} + \mathbf{K}_{xy}\right)y\widehat{\mathbf{T}}_y}{x} \\
&+ \widehat{\mathbf{T}}_y^T\left(-\omega^2\,\mathbf{M}_y + i\omega\mathbf{C}_y + \mathbf{K}_y\right)\widehat{\mathbf{T}}_y + \widehat{\mathbf{T}}_y^T\left(i\omega\mathbf{C}_{yxy} + \mathbf{K}_{yxy}\right)x\widehat{\mathbf{T}}_{xy} \\
&+ \frac{\widehat{\mathbf{T}}_y^T\left(i\omega\mathbf{C}_{my}^T + \mathbf{K}_{my}^T\right)}{y} + \frac{\widehat{\mathbf{T}}_y^T\left(i\omega\mathbf{C}_{xy}^T + \mathbf{K}_{xy}^T\right)x\widehat{\mathbf{T}}_x}{y} \\
&+ \widehat{\mathbf{T}}_{xy}^T\left(-\omega^2\,\mathbf{M}_{xy} + i\omega\mathbf{C}_{xy} + \mathbf{K}_{xyxy}\right)\widehat{\mathbf{T}}_{xy} \\
&+ \frac{\widehat{\mathbf{T}}_{xy}^T\left(i\omega\mathbf{C}_{yxy}^T + \mathbf{K}_{yxy}^T\right)\widehat{\mathbf{T}}_y}{x} + \frac{\widehat{\mathbf{T}}_{xy}^T\left(i\omega\mathbf{C}_{xxy}^T + \mathbf{K}_{xxy}^T\right)\widehat{\mathbf{T}}_x}{y} \\
&+ \frac{\widehat{\mathbf{T}}_{xy}^T\left(i\omega\mathbf{C}_{mxy}^T + \mathbf{K}_{mxy}^T\right)}{xy}
\end{aligned} \tag{15}
$$

Our goal is to determine the highest degree of $x$ and $y$ in the determinant of $\mathbf{D}$. Using the Leibniz formula to calculate the determinant, each term in the polynomial of det($\mathbf{D}$) consists of exactly one element from each row and column. Thus, in order to get the highest power of $x$ and $y$, in each row/column, we pick the term with the highest power of $x$ and $y$. Now it is noted that $\mathbf{D}_{11}$ and $\mathbf{D}_{21}$ do not contain any term with positive powers of $x$ or $y$. In general, dispersion analysis is performed in the following Brillouin zone cases:

- $x$ varies while $y$ is a constant number.
- $y$ varies while $x$ is a constant number.
- $y=x^a$ , $a$ can assume any real number.

In most of the dispersion curve analysis however, only the boundary of irreducible Brillouin zone is considered [35], meaning that ω is calculated for the first two lines and a single value of $a$ for the third line. In the example of a square lattice – such as the one in Fig. 2 - these lines are noted by Γ-X, X-M and M- Γ, in which the line M- Γ corresponds to the third line with $a$ = 1. In our representation, we can cover more of the Brillouin zone by picking more values for $a$. We investigate the degree of $x$ or $y$ on each of these lines separately.

### 3.1. First line: $x$ varies, $y$ is a constant number

On the first line we will show that the degree of $x$ in the det($\mathbf{D}$) is bounded above by:

$$Rank(\widehat{\mathbf{D}}_x) \tag{16}$$

in which,

$$\widehat{\mathbf{D}}_x = \begin{bmatrix}\mathbf{D}_{1x}\\ \mathbf{D}_{2x}\end{bmatrix}, \tag{17}$$

and

$$\mathbf{D}_{1x} = (i\omega\mathbf{C}_{ix} + \mathbf{K}_{ix})\widehat{\mathbf{T}}_x + \left(i\omega\mathbf{C}_{ixy} + \mathbf{K}_{ixy}\right)\widehat{\mathbf{T}}_{xy} \tag{18}$$

$$
\begin{aligned}
\mathbf{D}_{2x} = (i\omega\mathbf{C}_{mx} + \mathbf{K}_{mx})\widehat{\mathbf{T}}_x &+ \left(i\omega\mathbf{C}_{mxy} + \mathbf{K}_{mxy}\right)\widehat{\mathbf{T}}_{xy} + \widehat{\mathbf{T}}_y^T\left(i\omega\mathbf{C}_{yxy} + \mathbf{K}_{yxy}\right)\widehat{\mathbf{T}}_{xy} \\
&+ \widehat{\mathbf{T}}_y^T\left(i\omega\mathbf{C}_{xy}^T + \mathbf{K}_{xy}^T\right)\widehat{\mathbf{T}}_x
\end{aligned} \tag{19}
$$

Note that $\mathbf{D}_{1x}$ and $\mathbf{D}_{2x}$ consist of those terms in $\mathbf{D}_{12}$ and $\mathbf{D}_{22}$ with dependence on $x$.
Now for the sake of simplicity, it is safe to assume damping is zero whenever stiffness is zero, so without altering the rank of $\widehat{\mathbf{D}}_x$, we can simplify $\mathbf{D}_{1x}$ and $\mathbf{D}_{2x}$ as:

$$\mathbf{D}'_{1x} = \mathbf{K}_{ix}\widehat{\mathbf{T}}_x + \mathbf{K}_{ixy}\widehat{\mathbf{T}}_{xy} \tag{20}$$

$$\mathbf{D}'_{2x} = \mathbf{K}_{mx}\widehat{\mathbf{T}}_x + \mathbf{K}_{mxy}\widehat{\mathbf{T}}_{xy} + \widehat{\mathbf{T}}_y^T\mathbf{K}_{yxy}\widehat{\mathbf{T}}_{xy} + \widehat{\mathbf{T}}_y^T\mathbf{K}_{xy}^T\widehat{\mathbf{T}}_x \tag{21}$$

In order to show that on the line Γ-X the determinant of the matrix $\mathbf{D}$ is a function of $x$ with highest power bounded above by (16), we use the Leibniz formula. First we make an augmented matrix by matrix $\mathbf{D}$ next to $\widehat{\mathbf{D}}_x$:

$$\mathbf{D}_{aug} = \left[\begin{matrix}\mathbf{D}_{11} & \mathbf{D}_{12}\\ \mathbf{D}_{21} & \mathbf{D}_{22}\end{matrix}\middle|\begin{matrix}\mathbf{D}_{1x}\\ \mathbf{D}_{2x}\end{matrix}\right] \tag{22}$$

The determinant of $\mathbf{D}$ in (11) remains constant by any sequence of row operations. As such we can apply a sequence of row operations on $\mathbf{D}_{aug}$ to make $\widehat{\mathbf{D}}_x$ a row echelon matrix. Note that due to the relation between $\widehat{\mathbf{D}}_x$ and $\mathbf{D}_{12}$ and $\mathbf{D}_{22}$, making $\widehat{\mathbf{D}}_x$ a row echelon matrix, makes the second column of $\mathbf{D}$ a row echelon matrix in terms of $x$. This means that any row or column has at most one entry with $x$. Now, according to the Leibniz formula, any term in the determinant of $\mathbf{D}$ has exactly one element from each row and column. Since $\mathbf{D}_{12}$ and $\mathbf{D}_{22}$ do not contain any positive power of $x$, the highest power of $x$ in det($\mathbf{D}$) is achieved by choosing from each row the leading entry of $[\mathbf{D}_{11} \quad \mathbf{D}_{22}]^T$ with $x$. It can be shown [37] that the rank of a matrix equals the number of non-zero pivots of a row echelon form of that matrix. Consequently, the highest power of $x$ in det($\mathbf{D}$) equals to $Rank(\widehat{\mathbf{D}}_x)$. It should be noted here, that due to symmetry, both $x$ and $x^{-1}$ are roots of det($\mathbf{D}$)=0, thus we can multiply both sides by $x$ to the power of $Rank(\widehat{\mathbf{D}}_x)$ to obtain an equation in which the left hand side is a polynomial of $x$ with power of $2Rank(\widehat{\mathbf{D}}_x)$. To illustrate the proof, first we consider a simple matrix and then the example structure of Fig. 2.

In the following matrix,

$$\begin{bmatrix} 1 & 2 & x & 2x+2 & 2 \\ 4 & 5 & 2 & 0 & 3 \\ 2 & 2 & x+1 & x+3 & 1 \\ 2+\frac{1}{x} & 5+\frac{2}{x} & 3x+1+\frac{2}{x} & 3x+\frac{1}{x} & 5 \\ 1 & 2+\frac{3}{x} & 4 & 1 & 2+\frac{6}{x} \end{bmatrix} \tag{23}$$

We have

$$\mathbf{D}_{11} = \begin{bmatrix} 1 & 2 \\ 4 & 5 \\ 2 & 2 \end{bmatrix},\ \mathbf{D}_{12} = \begin{bmatrix} x & 2x+2 & 2 \\ 2 & 0 & 3 \\ x+1 & x+3 & 1 \end{bmatrix},$$
$$\mathbf{D}_{21} = \begin{bmatrix} 2+\frac{1}{x} & 5+\frac{2}{x} \\ 1 & 2+\frac{3}{x} \end{bmatrix}, \mathbf{D}_{22} = \begin{bmatrix} 3x+1+\frac{2}{x} & 3x+\frac{1}{x} & 5 \\ 4 & 1 & 2+\frac{6}{x} \end{bmatrix}. \tag{24}$$

Then $\mathbf{D}_{1x}$ and $\mathbf{D}_{2x}$ are:

$$\mathbf{D}_{1x} = \begin{bmatrix} 1 & 2 & 0 \\ 0 & 0 & 0 \\ 1 & 1 & 0 \end{bmatrix}, \mathbf{D}_{2x} = \begin{bmatrix} 3 & 3 & 0 \\ 0 & 0 & 0 \end{bmatrix} \tag{25}$$

Which makes $\mathbf{D}_{aug}$ in the form of:

$$\mathbf{D}_{aug} = \left[\begin{array}{ccccc|ccc} 1 & 2 & x & 2x+2 & 2 & 1 & 2 & 0 \\ 4 & 5 & 2 & 0 & 3 & 0 & 0 & 0 \\ 2 & 2 & x+1 & x+3 & 1 & 1 & 1 & 0 \\ 2+\frac{1}{x} & 5+\frac{2}{x} & 3x+1 & 3x & 5 & 3 & 3 & 0 \\ 1 & 2+\frac{3}{x} & 4 & 1 & 2 & 0 & 0 & 0 \end{array}\right] \tag{26}$$

After making the right matrix a row echelon matrix by row operations, we get the matrix:

$$\mathbf{D}_{aug} = \left[\begin{array}{ccccc|ccc} 3 & 2 & x+2 & 4 & 0 & 1 & 0 & 0 \\ -1 & 0 & -1 & x-1 & 1 & 0 & 1 & 0 \\ 4 & 5 & 2 & 0 & 3 & 0 & 0 & 0 \\ -4+\frac{1}{x} & -1+\frac{2}{x} & -2+\frac{2}{x} & -9+\frac{1}{x} & 2 & 0 & 0 & 0 \\ 1 & 2+\frac{3}{x} & 4 & 1 & 2+\frac{6}{x} & 0 & 0 & 0 \end{array}\right] \tag{27}$$

Evidently, in (27) the rank the right block matrix is two. It can also be verified easily that $Rank([\mathbf{D}_{1x} \quad \mathbf{D}_{2x}]^T)$ is also two. There are two nonzero columns on the right block matrix which correspond to two columns (third and fourth) on the left block matrix containing $x$. The determinant of matrix D is:

$$\det(\mathbf{D}) = \frac{5x^4 + 5x^4 + 131x^2 + 127x + 75}{x^2} \tag{28}$$

Multiplying both sides by $x^2$ we obtain an equation with $x$ to the power of four which agrees with that expected.

This relation between the rank of $\widehat{\mathbf{D}}_x$ and the $x$ power can also be investigated for the example structure of Fig. 2 with out of plane motion. Damping terms have been omitted for simplicity of notation. For this structure displacement vectors were defined in eq *(2)*. Stiffness, mass and linear transformation matrices are:

$$\mathbf{K}_i = \begin{bmatrix} \mathrm{K}_1+\mathrm{K}_4+\mathrm{K}_5 & -\mathrm{K}_1 & 0 & -\mathrm{K}_4 \\ -\mathrm{K}_1 & \mathrm{K}_1+\mathrm{K}_2+\mathrm{K}_6 & -\mathrm{K}_2 & 0 \\ 0 & -\mathrm{K}_2 & \mathrm{K}_2+\mathrm{K}_3+\mathrm{K}_7 & -\mathrm{K}_3 \\ -\mathrm{K}_4 & 0 & -\mathrm{K}_3 & \mathrm{K}_3+\mathrm{K}_4+\mathrm{K}_8 \end{bmatrix}$$

$$\mathbf{K}_{im} = \begin{bmatrix} 0 \\ 0 \\ 0 \\ -\mathrm{K}_8 \end{bmatrix}, \mathbf{K}_{ix} = \begin{bmatrix} 0 \\ 0 \\ -\mathrm{K}_7 \\ 0 \end{bmatrix}, \mathbf{K}_{iy} = \begin{bmatrix} -\mathrm{K}_5 \\ 0 \\ 0 \\ 0 \end{bmatrix}, \mathbf{K}_{ixy} = \begin{bmatrix} 0 \\ -\mathrm{K}_6 \\ 0 \\ 0 \end{bmatrix} \tag{29}$$

$$\mathbf{K}_{mx} = -\mathrm{K}_{10}, \mathbf{K}_{yxy} = -\mathrm{K}_{10}, \mathbf{K}_{xxy} = -\mathrm{K}_9, \mathbf{K}_{mxy} = 0$$

$$\mathbf{K}_{xy} = 0, \mathbf{K}_m = 0, \mathbf{K}_{xyxy} = 0, \mathbf{K}_x = 0, \mathbf{K}_y = 0$$

and:

$$
\mathbf{M}=\begin{bmatrix} m_1 & 0 & 0 & 0 & 0 & 0 & 0 & 0 \\ 0 & m_2 & 0 & 0 & 0 & 0 & 0 & 0 \\ 0 & 0 & m_3 & 0 & 0 & 0 & 0 & 0 \\ 0 & 0 & 0 & m_4 & 0 & 0 & 0 & 0 \\ 0 & 0 & 0 & 0 & m_5 & 0 & 0 & 0 \\ 0 & 0 & 0 & 0 & 0 & m_5 & 0 & 0 \\ 0 & 0 & 0 & 0 & 0 & 0 & m_5 & 0 \\ 0 & 0 & 0 & 0 & 0 & 0 & 0 & m_5 \end{bmatrix}, \mathbf{T}=\begin{bmatrix} \mathbf{I} & \mathbf{0} \\ \mathbf{0} & 1 \\ \mathbf{0} & x \\ \mathbf{0} & y \\ \mathbf{0} & xy \end{bmatrix}, \quad (30)
$$

in which $\mathbf{I}$ is a 4x4 matrix. Using eq. (20) and eq. (21), for this structure we have:

$$
\mathbf{D}_{1x}=\begin{bmatrix} 0 \\ 0 \\ -\mathrm{K}_7 \\ 0 \end{bmatrix}\times 1+\begin{bmatrix} 0 \\ -\mathrm{K}_6 \\ 0 \\ 0 \end{bmatrix}\times 1=\begin{bmatrix} 0 \\ -\mathrm{K}_6 \\ -\mathrm{K}_7 \\ 0 \end{bmatrix}, \mathbf{D}_{2x}=-\mathbf{2}\mathrm{K}_{10}-\mathrm{K}_9 \quad (31)
$$

Then:

$$
\widehat{\mathbf{D}}_x=\begin{bmatrix} 0 \\ -\mathrm{K}_6 \\ -\mathrm{K}_7 \\ 0 \\ -2\mathrm{K}_{10}-\mathrm{K}_9 \end{bmatrix} \quad (32)
$$

Since $\widehat{\mathbf{D}}_x$ is a column matrix, its rank is clearly 1. To verify this result, we find the equation det(**D**)=0. This equation, with the use of symbolic manipulator can be verified to be in the form of:

$$
\det(\mathbf{D})=\frac{x^2(A+By)+y^2(C+Dx)+Exy+Fx+Gy}{xy} \quad (33)
$$

where $A$, $B$, $C$, $D$, $E$, $F$ and $G$ are functions of spring constants, masses and ω. It should be noted that A and B have 234 terms, which makes them computationally costly to calculate and difficult to comprehend. On the other hand, the procedure of calculating rank of $\widehat{\mathbf{D}}_x$ is simple and does not require the handling of constants with many terms.

### 3.2. Second line: $y$ varies, $x$ is a constant number

For the second line, we borrow the same argument we used in the previous section. It can be verified that the $y$ degree in the det(**D**)=0 is bounded above by:

$$
Rank(\widehat{\mathbf{D}}_y) \quad (34)
$$

In which,

$$
\widehat{\mathbf{D}}_y=\begin{bmatrix} \mathbf{D}_{1y} \\ \mathbf{D}_{2y} \end{bmatrix}, \quad (35)
$$

Similar to the previous case, for the sake of simplicity, we assume zero damping whenever stiffness is zero, so we can shorten $\mathbf{D}_{1y}$ and $\mathbf{D}_{2y}$ terms as:

$$
\mathbf{D}'_{1y}=\mathbf{K}_{iy}\widehat{\mathbf{T}}_y+\mathbf{K}_{ixy}\widehat{\mathbf{T}}_{xy} \quad (36)
$$

$$
\mathbf{D}'_{2y}=\mathbf{K}_{my}\widehat{\mathbf{T}}_y+\mathbf{K}_{mxy}\widehat{\mathbf{T}}_{xy}+\widehat{\mathbf{T}}_x^T\mathbf{K}_{xxy}\widehat{\mathbf{T}}_{xy}+\widehat{\mathbf{T}}_x^T\mathbf{K}_{xy}\widehat{\mathbf{T}}_y \quad (37)
$$

### 3.3. Third line: $y = x^a$

On the third line, $y = x^a$. Here, for the ease of explanation, we first consider $y = x$ before moving on to the more general case. The relation $y = x$ makes all the $xy$ terms in matrix $\mathbf{D}$ into power two terms of $x$; namely $x^2$. Therefore, in estimating the highest power of $x$ in the det($\mathbf{D}$), we count the number of $xy$, and $x$ or $y$, in a different fashion. On the line $y = x$, the variable $y$ is represented by $x$. We define two matrices:

$$\widehat{\mathbf{D}}_{ax} = \begin{bmatrix} \mathbf{D}_{1ax} \\ \mathbf{D}_{2ax} \end{bmatrix}, \tag{38}$$

In which

$$\mathbf{D}_{1ax} = (i\omega\mathbf{C}_{ix} + \mathbf{K}_{ix})\widehat{\mathbf{T}}_x + \left(i\omega\mathbf{C}_{iy} + \mathbf{K}_{iy}\right)\widehat{\mathbf{T}}_y \tag{39}$$

$$\mathbf{D}_{2ax} = (i\omega\mathbf{C}_{mx} + \mathbf{K}_{mx})\widehat{\mathbf{T}}_x + \left(i\omega\mathbf{C}_{my} + \mathbf{K}_{my}\right)\widehat{\mathbf{T}}_y + \widehat{\mathbf{T}}_x^T\left(i\omega\mathbf{C}_{xxy} + \mathbf{K}_{xxy}\right)\widehat{\mathbf{T}}_{xy} + \widehat{\mathbf{T}}_y^T\left(i\omega\mathbf{C}_{yxy} + \mathbf{K}_{yxy}\right)\widehat{\mathbf{T}}_{xy} \tag{40}$$

These are the terms in eq.(13) and (15) which include $x$ or $y$. The other matrix consists of the terms containing $xy$:

$$\widehat{\mathbf{D}}_{xy} = \begin{bmatrix} \mathbf{D}_{1xy} \\ \mathbf{D}_{2xy} \end{bmatrix}, \tag{41}$$

In which

$$\mathbf{D}_{1xy} = \left(i\omega\mathbf{C}_{ixy} + \mathbf{K}_{ixy}\right)\widehat{\mathbf{T}}_{xy} \tag{42}$$

$$\mathbf{D}_{2xy} = \left(i\omega\mathbf{C}_{mxy} + \mathbf{K}_{mxy}\right)\widehat{\mathbf{T}}_{xy} \tag{43}$$

Then we show that the power of $x$ in the det($\mathbf{D}$) is bounded above by:

$$2 \times Rank\left(\widehat{\mathbf{D}}_{xy}\right) + (Rank(\widehat{\mathbf{D}}_{xy} + \widehat{\mathbf{D}}_{ax}) - Rank(\widehat{\mathbf{D}}_{xy})) \tag{44}$$

Which can be simplified to:

$$Rank\left(\widehat{\mathbf{D}}_{xy}\right) + Rank(\widehat{\mathbf{D}}_{xy} + \widehat{\mathbf{D}}_{ax}) \tag{45}$$

The rationale behind (45) is explained through an example. Consider Matrix $\mathbf{A}$ as follows:

$$\mathbf{A} = \begin{bmatrix} x^2 + x & 0 & 0 & 0 & 0 \\ x^2 & 1 & 0 & x^2 & 0 \\ x^2 & 0 & x & 0 & 0 \\ x^2 & 0 & x & x^2 & x^2 \\ x & 0 & 0 & x^2 & x \end{bmatrix} \tag{46}$$

Considering only the variables $x$ and $x^2$, matrix $\mathbf{A}$ has rank four. Matrices $\mathbf{A}_x$ and $\mathbf{A}_{xy}$ consist of those elements with $x$ and $x^2$, respectively. They are:

$$\mathbf{A}_x = \begin{bmatrix} 1 & 0 & 0 & 0 & 0 \\ 0 & 0 & 0 & 0 & 0 \\ 0 & 0 & 1 & 0 & 0 \\ 0 & 0 & 1 & 0 & 0 \\ 1 & 0 & 0 & 0 & 1 \end{bmatrix}, \mathbf{A}_{xy} = \begin{bmatrix} 1 & 0 & 0 & 0 & 0 \\ 1 & 0 & 0 & 1 & 0 \\ 1 & 0 & 0 & 0 & 0 \\ 1 & 0 & 0 & 1 & 1 \\ 0 & 0 & 0 & 1 & 0 \end{bmatrix} \tag{47}$$

For these matrices we have:

$$Rank(\mathbf{A}_x) = 3, Rank\left(\mathbf{A}_{xy}\right) = 3, Rank\left(\mathbf{A}_x + \mathbf{A}_{xy}\right) = Rank(\mathbf{A}) = 4 \tag{48}$$

The determinant of $\mathbf{A}$ can be verified to be:

$$det(\mathbf{A}) = (x^2 + x)x^4(1 - x) \tag{49}$$

which is a polynomial of degree 7. Also:

$$Rank(\mathbf{A}_{xy}) + Rank(\mathbf{A}_x + \mathbf{A}_{xy}) = Rank(\mathbf{A}_{xy}) + Rank(\mathbf{A}) = 3 + 4 = 7 \tag{50}$$

To illustrate the reason behind (45), we first make the matrix $\mathbf{A}$ into a row echelon matrix in terms of $x^2$. This process uses row operations and therefore does not alter the det($\mathbf{A}$). After these row operations matrix $\mathbf{A}$ takes the form:

$$\mathbf{A} = \begin{bmatrix} x^2 + x & 0 & 0 & 0 & 0 \\ -x & 1 & 0 & x^2 & 0 \\ 0 & -1 & x & 0 & x^2 \\ -x & 0 & x & 0 & 0 \\ 2x & -1 & 0 & 0 & x \end{bmatrix} \tag{51}$$

A column operation does not alter the determinant. We perform column operations such that the columns containing pivots are arranged next to each other, *i.e.*,

$$\mathbf{A} = \begin{bmatrix} x^2 + x & 0 & 0 & 0 & 0 \\ -x & x^2 & 0 & 1 & 0 \\ 0 & 0 & x^2 & -1 & x \\ -x & 0 & 0 & 0 & x \\ 2x & 0 & x & -1 & 0 \end{bmatrix} \tag{52}$$

In calculating the determinant of (52) by the Leibniz formula, the highest degree of $x$ is achieved by picking $x^2$ from the first three columns and rows. We perform row operations further to make the corner 2 x 2 matrix

$$\begin{bmatrix} 0 & x \\ -1 & 0 \end{bmatrix}, \tag{53}$$

a row echelon matrix in $x$. This is not necessary in this case since it is already in row echelon form. Note that the number of linearly independent rows in this 2×2 matrix equals to:

$$Rank(\mathbf{A}_x + \mathbf{A}_{xy}) - Rank(\mathbf{A}_{xy}) = 4 - 3 = 1 \tag{54}$$

To further clarify eq. (44), if an element in matrix $\mathbf{D}$ contain both $x$ and $x^2$ , the highest power is defined $x^2$ . Also the row echelon operation should be done on the matrix $\mathbf{D}$ as a whole.

For the general case of $y = x^a$ , we consider cases for which a>1. For the situations in which a<1 we can simply swap the role of $x$ and $y$. By following similar procedure as the previous section with some modification, we have:

$$\widehat{\mathbf{D}}_{ax} = \begin{bmatrix} \mathbf{D}_{1ax} \\ \mathbf{D}_{2ax} \end{bmatrix}, \widehat{\mathbf{D}}_{ay} = \begin{bmatrix} \mathbf{D}_{1ay} \\ \mathbf{D}_{2ay} \end{bmatrix}, \ \widehat{\mathbf{D}}_{y/x} = \begin{bmatrix} \mathbf{0} \\ \mathbf{D}_{y/x} \end{bmatrix}, \widehat{\mathbf{D}}_{xy} = \begin{bmatrix} \mathbf{D}_{1xy} \\ \mathbf{D}_{2xy} \end{bmatrix} \tag{55}$$

In which,

$$\mathbf{D}_{1ax} = (i\omega \mathbf{C}_{ix} + \mathbf{K}_{ix})\widehat{\mathbf{T}}_x \tag{56}$$

$$\mathbf{D}_{2ax} = (i\omega \mathbf{C}_{mx} + \mathbf{K}_{mx})\widehat{\mathbf{T}}_x + \widehat{\mathbf{T}}_y^T(i\omega \mathbf{C}_{yxy} + \mathbf{K}_{yxy})\widehat{\mathbf{T}}_{xy} \tag{57}$$

$$\mathbf{D}_{1ay} = (i\omega \mathbf{C}_{iy} + \mathbf{K}_{iy})\widehat{\mathbf{T}}_y \tag{58}$$

$$\mathbf{D}_{2ay} = (i\omega \mathbf{C}_{my} + \mathbf{K}_{my})\widehat{\mathbf{T}}_y + \widehat{\mathbf{T}}_x^T(i\omega \mathbf{C}_{xxy} + \mathbf{K}_{xxy})\widehat{\mathbf{T}}_{xy} \tag{59}$$

$$\mathbf{D}_{y/x} = \widehat{\mathbf{T}}_x^T(i\omega \mathbf{C}_{xy} + \mathbf{K}_{xy})\widehat{\mathbf{T}}_y \tag{60}$$

$$\mathbf{D}_{1xy} = \left(i\omega\mathbf{C}_{ixy} + \mathbf{K}_{ixy}\right)\widehat{\mathbf{T}}_{xy} \tag{61}$$

$$\mathbf{D}_{2xy} = \left(i\omega\mathbf{C}_{mxy} + \mathbf{K}_{mxy}\right)\widehat{\mathbf{T}}_{xy} \tag{62}$$

$\mathbf{D}_{y/x}$ are the terms in eq (15) containing $y/x$. The power of $x$ in $y/x$ is higher or lower than $x$ depending on the value of a. We consider both cases; if 2≥a>1, power of $x$ in the det($\mathbf{D}$) is bounded above by:

$$\begin{aligned}(a+1)\times Rank\big(\widehat{\mathbf{D}}_{xy}\big) &+ \Big(a\times Rank\big(\widehat{\mathbf{D}}_{ay}+\widehat{\mathbf{D}}_{xy}\big) - Rank\big(\widehat{\mathbf{D}}_{xy}\big)\Big)\\ &+ \Big(Rank\big(\widehat{\mathbf{D}}_{ax}+\widehat{\mathbf{D}}_{ay}+\widehat{\mathbf{D}}_{xy}\big) - Rank\big(\widehat{\mathbf{D}}_{ay}+\widehat{\mathbf{D}}_{xy}\big)\Big)\\ &+ (a-1)\Big(Rank\big(\widehat{\mathbf{D}}_{y/x}+\widehat{\mathbf{D}}_{ax}+\widehat{\mathbf{D}}_{ay}+\widehat{\mathbf{D}}_{xy}\big)\\ &- Rank\big(\widehat{\mathbf{D}}_{ax}+\widehat{\mathbf{D}}_{ay}+\widehat{\mathbf{D}}_{xy}\big)\Big)\end{aligned}$$

Which can be simplified to:

$$\begin{aligned}&a\times Rank\big(\widehat{\mathbf{D}}_{xy}\big) + (a-1)\times Rank\big(\widehat{\mathbf{D}}_{ay}+\widehat{\mathbf{D}}_{xy}\big) +\\ &(2-a)Rank\big(\widehat{\mathbf{D}}_{ax}+\widehat{\mathbf{D}}_{ay}+\widehat{\mathbf{D}}_{xy}\big) + (a-1)Rank\big(\widehat{\mathbf{D}}_{y/x}+\widehat{\mathbf{D}}_{ax}+\widehat{\mathbf{D}}_{ay}+\widehat{\mathbf{D}}_{xy}\big)\end{aligned} \tag{63}$$

For the case of a>2, power of $x$ in the det($\mathbf{D}$) is bounded above by:

$$\begin{aligned}a\times Rank\big(\widehat{\mathbf{D}}_{xy}\big) &+ 1\times Rank\big(\widehat{\mathbf{D}}_{ay}+\widehat{\mathbf{D}}_{xy}\big)\\ &+ (a-2)Rank\big(\widehat{\mathbf{D}}_{y/x}+\widehat{\mathbf{D}}_{ay}+\widehat{\mathbf{D}}_{xy}\big)\\ &+ Rank\big(\widehat{\mathbf{D}}_{y/x}+\widehat{\mathbf{D}}_{ax}+\widehat{\mathbf{D}}_{ay}+\widehat{\mathbf{D}}_{xy}\big)\end{aligned} \tag{64}$$

In the next section, we investigate an example application of this theorem.

## 4. INVESTIGATING PHONON DISPERSION CURVES IN BORON NITRIDE

Boron Nitride is a chemical compound which is produced synthetically from boron acid or diboron trioxide among other methods. It consists of equal numbers of boron and nitrogen atoms and exists in different crystalline structures. Very similar to carbon compounds, it has a hexagonal form (h-BN) like graphite, and a cubic form (c- BN) like diamond. Among other applications, the hexagonal form is used as lubricant. Unlike graphite, h-BN can perform as a lubricant without molecules of air or water trapped between its layers. Compared to diamond, c-BN has superior thermal and chemical stability. Similar to carbon nanotubes, there exists BN nanotubes. They have been theoretically predicted [38] and experimentally verified to exist [39]. In this section, as an example application, we use formulas developed in the previous section to show that phonon dispersion curves of h-BN cannot be reproduced by considering only the nearest neighbor interactions. The theoretical phonon dispersion curves of h-BN are obtained by *ab initio* calculation. Each unit cell of h-BN consists of 18 atoms in three parallel hexagons, three nitrogen and three boron atoms in each hexagons (see Fig. 4). Each atom can move in three directions, so in our formalism we have:

$$\tilde{\mathbf{q}} = \begin{bmatrix} \mathbf{q_1} \\ \mathbf{q_2} \\ \mathbf{q_3} \\ \mathbf{q_4} \end{bmatrix}, \tilde{\mathbf{q}}_\mathbf{x} = \begin{bmatrix} \mathbf{q_5} \\ \mathbf{q_6} \end{bmatrix}, \tilde{\mathbf{q}}_\mathbf{y} = \begin{bmatrix} \mathbf{q_7} \\ \mathbf{q_8} \end{bmatrix}, \tilde{\mathbf{q}}_\mathbf{z} = \begin{bmatrix} \mathbf{q_9} \\ \mathbf{q_{10}} \end{bmatrix}, \tilde{\mathbf{q}}_\mathbf{xy} = \begin{bmatrix} \mathbf{q_{11}} \\ \mathbf{q_{12}} \\ \mathbf{q_{13}} \\ \mathbf{q_{14}} \end{bmatrix}, \tag{65}$$

$$\tilde{\mathbf{q}}_\mathbf{xz} = [\mathbf{q_{15}}], \tilde{\mathbf{q}}_\mathbf{yz} = [\mathbf{q_{16}}], \tilde{\mathbf{q}}_\mathbf{xyz} = \begin{bmatrix} \mathbf{q_{17}} \\ \mathbf{q_{18}} \end{bmatrix},$$

in which $\mathbf{q_j}$ ; j = 1..18 are vectors of length 3. There is no internal masses, so we omitted $\mathbf{q}_i$. For the transformation matrices, we have:

$$\mathbf{T}_x = x\begin{bmatrix} \mathbf{I} & \mathbf{0} & \mathbf{0} & \mathbf{0} \\ \mathbf{0} & \mathbf{0} & \mathbf{I} & \mathbf{0} \end{bmatrix}, \mathbf{T}_y = y\begin{bmatrix} \mathbf{I} & \mathbf{0} & \mathbf{0} & \mathbf{0} \\ \mathbf{0} & \mathbf{0} & \mathbf{I} & \mathbf{0} \end{bmatrix}, \mathbf{T}_z = z\begin{bmatrix} \mathbf{I} & \mathbf{0} & \mathbf{0} & \mathbf{0} \\ \mathbf{0} & \mathbf{I} & \mathbf{0} & \mathbf{0} \end{bmatrix}$$

$$\mathbf{T}_{xz} = xz[\mathbf{I} \quad \mathbf{0} \quad \mathbf{0} \quad \mathbf{0}], \mathbf{T}_{yz} = yz[\mathbf{0} \quad \mathbf{I} \quad \mathbf{0} \quad \mathbf{0}], \mathbf{T}_{xy} = xy\begin{bmatrix} \mathbf{I} & \mathbf{0} & \mathbf{0} & \mathbf{0} \\ \mathbf{0} & \mathbf{I} & \mathbf{0} & \mathbf{0} \\ \mathbf{0} & \mathbf{0} & \mathbf{I} & \mathbf{0} \\ \mathbf{0} & \mathbf{0} & \mathbf{0} & \mathbf{I} \end{bmatrix} \tag{66}$$

$$\mathbf{T}_{xyz} = xyz\begin{bmatrix} \mathbf{I} & \mathbf{0} & \mathbf{0} & \mathbf{0} \\ \mathbf{0} & \mathbf{I} & \mathbf{0} & \mathbf{0} \end{bmatrix}$$

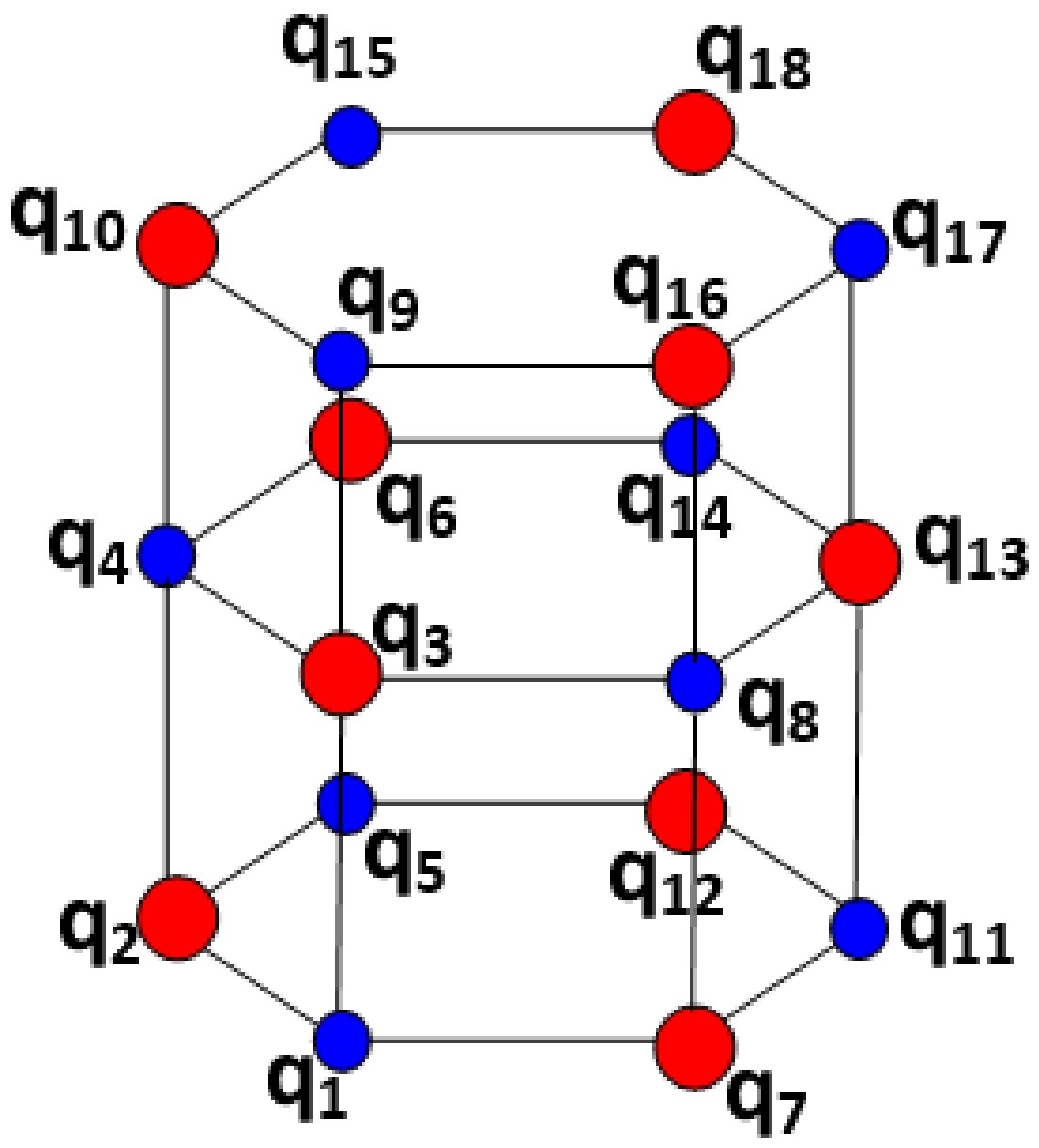

*Figure 4: Boron nitride in its hexagonal crystalline form (h-BN) and the unit cell displacements*

This crystal is three dimensional with z direction orthogonal to the hexagonal plane. There is no internal mass making $\mathbf{D'}_{1x}$ to be zero. $\mathbf{D'}_{2x}$ takes the form:

$$\begin{aligned} \mathbf{D'}_{2x} = {} & \mathbf{K}_{mx}\hat{\mathbf{T}}_x + \mathbf{K}_{mxy}\hat{\mathbf{T}}_{xy} + \mathbf{K}_{mxz}\hat{\mathbf{T}}_{xz} + \mathbf{K}_{mxyz}\hat{\mathbf{T}}_{xyz} \\ & + \hat{\mathbf{T}}_y^T\left(\mathbf{K}_{xy}^T\hat{\mathbf{T}}_x + \mathbf{K}_{yxy}\hat{\mathbf{T}}_{xy} + \mathbf{K}_{yxz}\hat{\mathbf{T}}_{xz} + \mathbf{K}_{yxyz}\hat{\mathbf{T}}_{xyz}\right) \\ & + \hat{\mathbf{T}}_z^T\left(\mathbf{K}_{xz}^T\hat{\mathbf{T}}_x + \mathbf{K}_{zxy}\hat{\mathbf{T}}_{xy} + \mathbf{K}_{zxz}\hat{\mathbf{T}}_{xz} + \mathbf{K}_{zxyz}\hat{\mathbf{T}}_{xyz}\right) \\ & + \hat{\mathbf{T}}_{yz}^T(\mathbf{K}_{xyz}^T\hat{\mathbf{T}}_x + \mathbf{K}_{xyyz}^T\hat{\mathbf{T}}_{xy} + \mathbf{K}_{xzyz}^T\hat{\mathbf{T}}_{xz} + \mathbf{K}_{yzxyz}\hat{\mathbf{T}}_{xyz}) \end{aligned} \tag{67}$$

Considering only the nearest neighbor interaction, most of the terms in (67) would be zero. After simple calculation, (67) takes the form:

$$\mathbf{D'}_{2x} = \begin{bmatrix} \mathbf{0} & \mathbf{0} & \mathbf{0} & \mathbf{0} \\ \mathbf{K}_\mathrm{A} & \mathbf{0} & \mathbf{0} & \mathbf{0} \\ \mathbf{0} & \mathbf{0} & \mathbf{0} & \mathbf{0} \\ \mathbf{0} & \mathbf{0} & \mathbf{K}_\mathrm{B} & \mathbf{0} \end{bmatrix} \tag{68}$$

in which $\mathbf{K}_\mathrm{A}$ and $\mathbf{K}_\mathrm{B}$ are some 3×3 matrices. Consequently:

$$Rank(\hat{\mathbf{D}}_x) = Rank(\mathbf{D'}_{2x}) = 6, \tag{69}$$

This means for each ω there would be no more than 6 wavevectors when we consider only the nearest neighbor interaction. However, as it is evident in Fig. 5 depicting phonon dispersion curve calculated by Wang et al [40], there is a range of ω's for which there corresponds more than six wavevectors. This indicates nearest neighbor interactions beyond the first neighbors are needed to model dispersion in this crystalline structures.

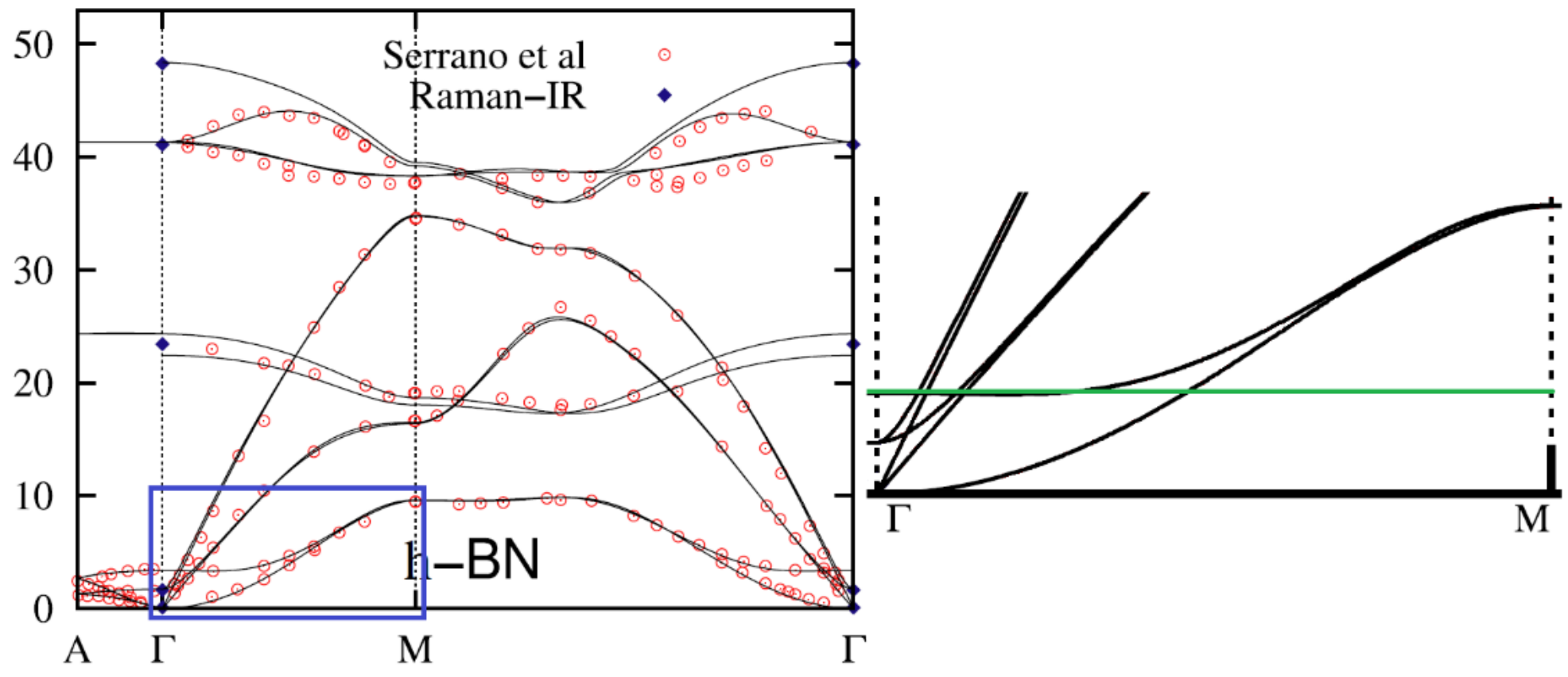

*Figure 5: Phonon dispersion curves of h-BN calculated by Wang et. al. [40] (solid lines) and X-Ray scattering (circles). The close-up of the blue box is depicted on the right. Green line shows an example value of ω for which we have more than 6 wavevectors.*

## 5. CONCLUDING REMARKS

In this paper, we studied the number of wavevectors for each frequency. We divided the boundary of irreducible Brillouin zone to different cases and investigated them separately. We provided a formula to calculate the upper bound for the maximum number of wavevectors for each frequency in a general periodic structure which might include damping. As an example application of the developed theory, we investigated phonon dispersion curves in h-BN to show that to model dispersion curve for this crystal, interactions beyond the first neighbor is needed. Presented theorem in this paper, can be thought of as a step toward designing periodic structures in which interactions is not limited to the closest neighbors.